\documentclass{appolb}
\usepackage{epsfig}

\newcommand{\sNN}{{{$\sqrt{s_{_{{NN}}}}$}}}

\newcommand{\gev}{\mbox{$\mathrm{GeV}$}}

\newcommand{\gevc}{\mbox{${\mathrm{GeV/}}c$}}

\newcommand{\KV}{{\mbox{$\kappa \sigma^{2}$}}}
\newcommand{\SD}{{\mbox{$S \sigma$}}}

\begin{document}
\title{Search for the QCD Critical Point by Higher Moments of
Net-proton Multiplicity Distributions at RHIC
\thanks{Presented at the conference `Strangeness in Quark Matter 2011`, Cracow, Poland,
September 18-24, 2011}%
}
\author{Xiaofeng Luo (for the STAR Collaboration)
\address{Institute of Particle Physics, Central China Normal University, Wuhan 430079, China}
\address{Key Laboratory of Quark and Lepton Physics (Central China Normal University), Ministry of Education, Wuhan 430079,
China}}
\maketitle
\begin{abstract}
Higher moments of net-proton multiplicity distributions are applied
to search for the QCD critical point. In this paper, we will present
measurements for kurtosis ($\kappa$), skewness ($S$) and variance
($\sigma^2$) of net-proton multiplicity distributions at the
mid-rapidity ($|y|<0.5$) and transverse momentum range
$0.4<p_{T}<0.8$ {\gevc} for Au+Au collisions at {\sNN} = 7.7, 11.5,
39, 62.4 and 200 GeV, Cu+Cu collisions at {\sNN} = 22.4, 62.4 and
200 GeV, $d$+Au collisions at {\sNN} = 200 GeV and $p$+$p$
collisions at {\sNN} = 62.4 and 200 GeV. The moment products $\kappa
\sigma^2$ and $S \sigma$ of net-proton distributions, which are
related to volume independent baryon number susceptibility ratios,
are consistent with Lattice QCD and Hadron Resonance Gas (HRG) model
calculations at high energies ({\sNN} = 62.4 and 200 GeV).
Deviations of $\kappa \sigma^2$ and $S\sigma$ for the Au+Au
collisions at low energies ({\sNN} = 7.7, 11.5 and 39 GeV) from HRG
model calculations are also observed.
\end{abstract}

\section{Introduction}
The main goal of Beam Energy Scan (BES) program~\cite{bes} at the
Relativistic Heavy Ion Collider (RHIC) is to study the phase
structure~\cite{science}, such as map the QCD phase boundary and
search for the QCD critical point~\cite{qcp}, of the QCD matter
created in heavy ion collision. By tuning the colliding energies of
two nuclei from {\sNN\ = 200 GeV} to {\sNN\ = 7.7 GeV}, we can
access various region of the QCD phase diagram. Higher moments
(variance ($\sigma^{2}$), skewness ($S$), kurtosis ($\kappa$) {\it
etc.}) of conserved quantities, such as net-baryon, net-charge and
net-strangeness, multiplicity distributions are very sensitive to
the correlation length~\cite{qcp_signal,ratioCumulant} and can be
directly connected to the corresponding thermodynamic
susceptibilities in Lattice QCD~\cite{Lattice,MCheng2009} and Hadron
Resonance Gas (HRG) model~\cite{HRG}. As the volume of the system is
hard to determine, the susceptibility ratio, such as
$\chi^{(4)}_{B}$/$\chi^{(2)}_{B}$ and
$\chi^{(3)}_{B}$/$\chi^{(2)}_{B}$, are used to compare with the
experimental data as $\kappa
\sigma^2=\chi^{(4)}_{B}$/$\chi^{(2)}_{B}$ and $S
\sigma=\chi^{(3)}_{B}$/$\chi^{(2)}_{B}$. Theoretical calculations
demonstrate that the experimental measurable net-proton (proton
number minus anti-proton number) number fluctuations can effectively
reflect the fluctuations of the net-baryon number~\cite{Hatta}.
Higher moments analysis opens a completely new domain and provides
quantitative method for probing the bulk properties of the hot dense
nuclear matter~\cite{science}.
\section{Observables}
Experimentally, we measure net-proton number event-by-event wise,
$N_{p-\bar{p}}=N_{p}-N_{\bar{p}}$, which is proton number minus
antiproton number. In the following, we use $N$ to represent the
net-proton number $N_{p-\bar{p}}$ in one event. The average value
over whole event ensemble is denoted by $ \mu=<N>$, where the single
angle brackets are used to indicate ensemble average of an
event-by-event distributions. The deviation of $N$ from its mean
value are defined by
\begin{equation}
  \delta N=N-<N>=N- \mu.
\end{equation}
The $r^{\rm th}$ order central moments are defined as
\begin{eqnarray}
  \mu _{r}  =  < (\delta N)^r  >,
  \mu _1  = 0.
\end{eqnarray}
Then, we can define various order cumulants of event-by-event
distributions as
\begin{eqnarray}
  C_1  &=&  \mu, C_2  =  \mu _2 , C_3  =  \mu _3,  \\
 C_n(n>3)  &=&  \mu _n  - \sum\limits_{m = 2}^{n - 2}
{\left(
\begin{array}{l}
 n - 1 \\
 m - 1 \\
 \end{array} \right) C_m }  \mu _{n - m}.
\end{eqnarray}
Once we have the definition of cumulants, various moments can be
denoted as
\begin{eqnarray}
 M=C_{1},\sigma^{2}= C_{2}, S=\frac{
C_{3}}{( C_{2})^{3/2}},\kappa=\frac{ C_{4}}{( C_{2})^{2}}.
\end{eqnarray}
Then, the moments product $\kappa\sigma^{2}$ and $S \sigma$ can be
expressed in term of cumulant ratio
\begin{eqnarray}
\kappa \sigma^{2}=\frac{ C_{4}}{ C_{2}},  S \sigma=\frac{ C_{3}}{
C_{2}}.
\end{eqnarray}

\section{Background Effects} 
\begin{figure}[htb]
 \hspace{-2cm}
\begin{minipage}[t]{0.6\linewidth}
\centering \vspace{0pt}
    \includegraphics[scale=0.35]{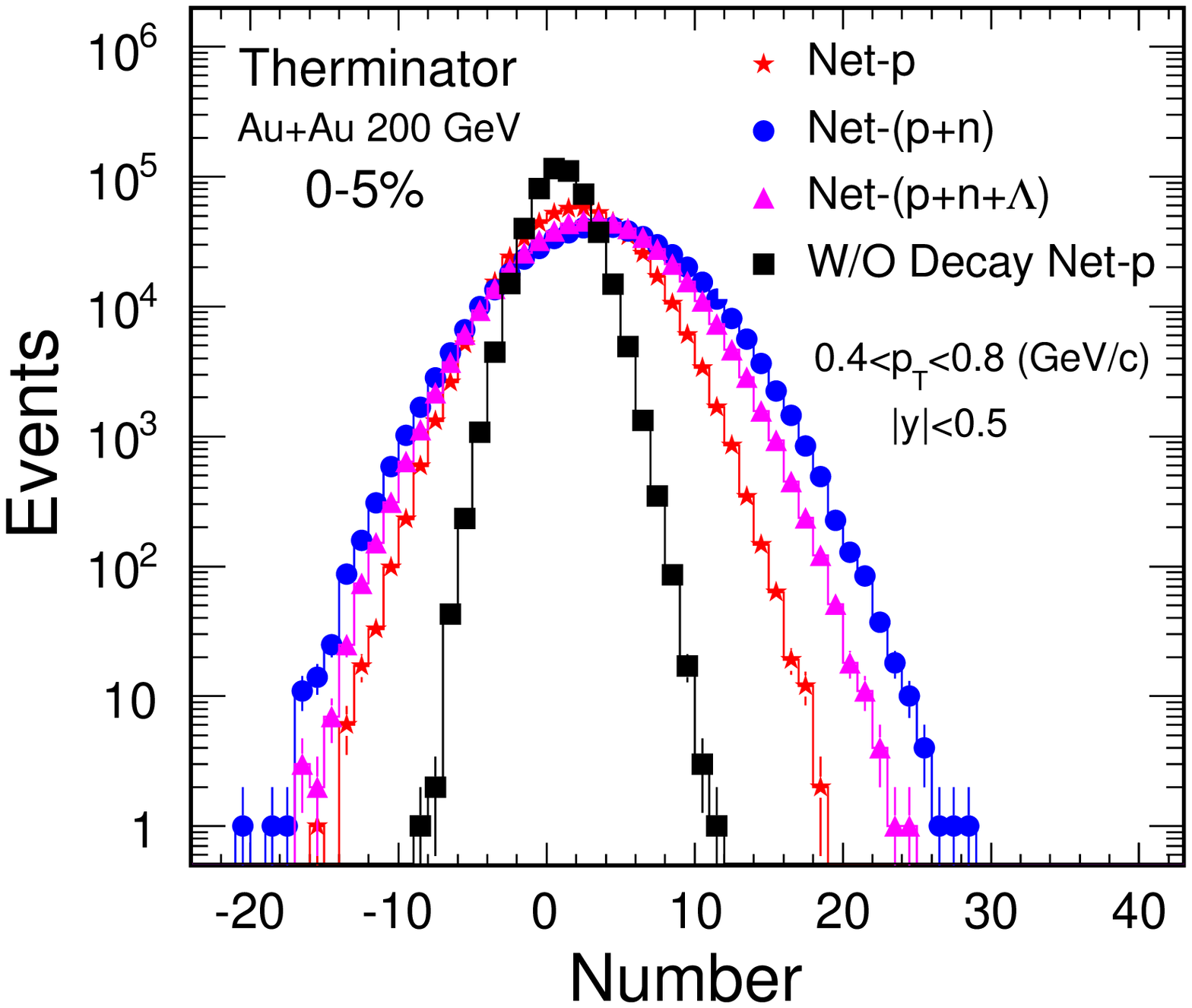}
   \caption{Event-by-event number distributions for net-proton,
net-(proton+neutron), net-(proton+neutron+$\Lambda$) and net-proton
without resonance decay for 0-5\% most central Au+Au collisions at
{\sNN} = 200 GeV from Therminator model.}
\label{fig:therminator_dis}
  \end{minipage}%
  \hspace{0.3in}
  \begin{minipage}[t]{0.6\linewidth}
  \centering \vspace{0pt}
   \includegraphics[scale=0.27]{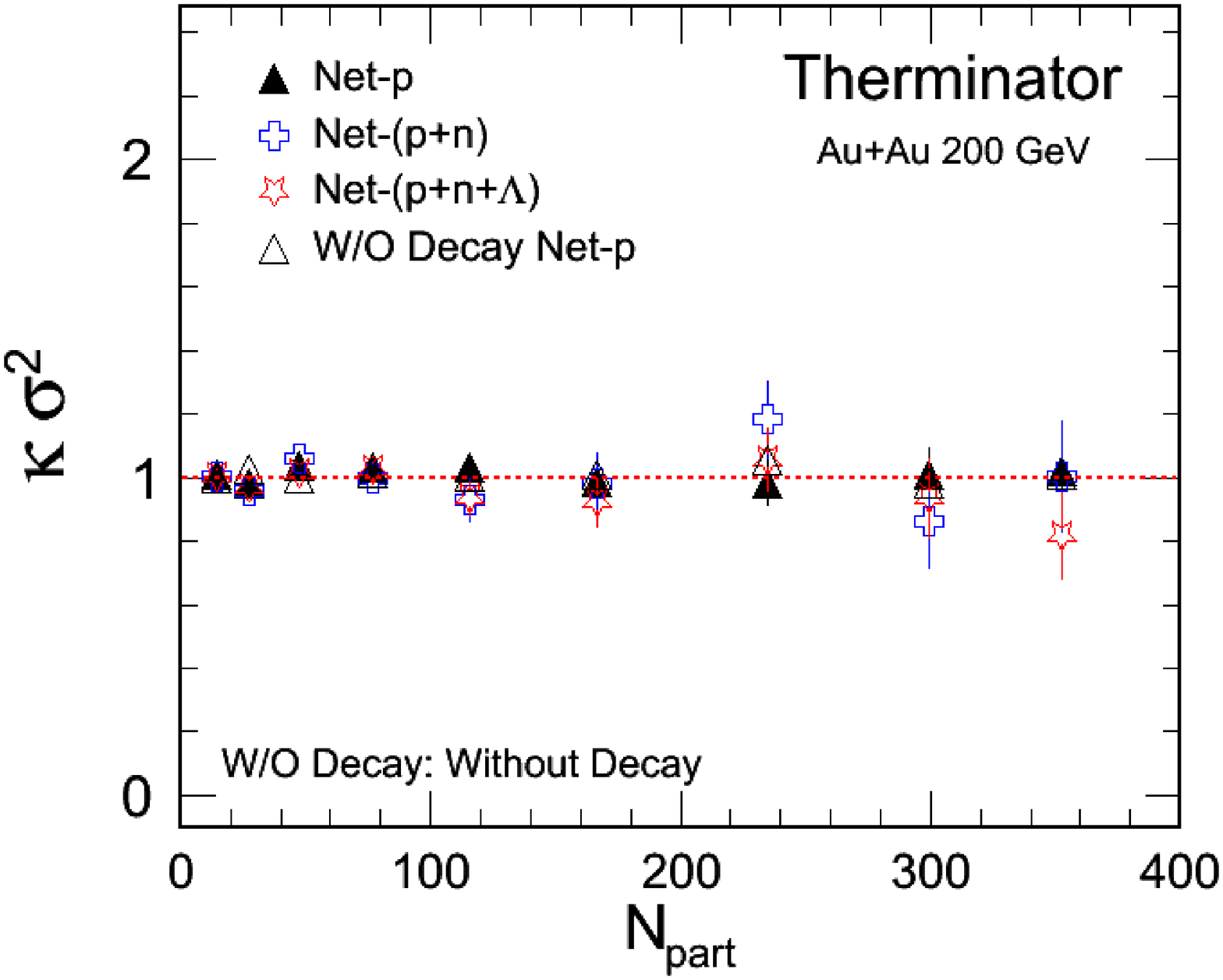}
    \caption[]
    {Centrality dependence of {\KV} of number distributions of four cases for Au+Au collisions at {\sNN} =200
{\gev} from Therminator model.} \label{fig:KV_therminator}
  \end{minipage} %
\end{figure}

Therminator model~\cite{thermal} was applied to study the
resonance decay effect, which is a background effect for higher
moment analysis, and also to check whether the net-proton
fluctuations can reflect the net-baryon number fluctuations. In Fig.
\ref{fig:therminator_dis}, we show the event-by-event number
distributions of Au+Au 0-5\% most central collisions at {\sNN} = 200
GeV from Therminator calculations for four cases. Fig.
\ref{fig:therminator_dis} demonstrates that the distribution for the
net-proton with resonance decay is wider than the net-proton
distributions without decay. By excluding the $\Lambda$ decay
nucleon, the net-(proton+neutron+$\Lambda$) has narrower
distribution than the net-(proton+neutron) distributions. Fig.
\ref{fig:KV_therminator} shows centrality dependence of {\KV} of
number distributions. The results for the four cases are consistent
with each other within errors, which indicate the effects of
resonance decay are small and the net-proton fluctuations can
reflect the net-baryon fluctuations. The statistical errors are
evaluated by Delta theorem method~\cite{Delta_theory}.

\section{Experimental Method}
The data presented in this proceedings are obtained using the
Solenoidal Tracker at RHIC (STAR). The main subsystem used in this
analysis is a large, uniform acceptance cylindrical Time Projection
Chamber (TPC) covering a pseudo-rapidity range of $|\eta|<1$ and
full azimuthal coverage. To ensure the purity and similar
efficiency, the protons and anti-protons are identified with the
ionization energy loss ($dE/dx$) measured by the TPC of STAR
detector within $0.4<p_{T}<0.8$ {\gevc} and mid-rapidity
($|y|<0.5$). Centralities are determined by the uncorrected charged
particle multiplicities ($dN_{ch}/d\eta$) within pseudo-rapidity
$|\eta|<0.5$ measured by the TPC and the centrality bin width
correction is used to eliminate volume fluctuations~\cite{WWND2011}.
By comparing measured $dN_{ch}/d\eta$ with the Monto Carlo Glauber
model results, we can obtain the average number of participant
($N_{\rm part}$) for each centrality.
\section{Results}
In this section, we present beam energy and system size dependence
of various moments ($M, \sigma, S, \kappa$) and moment products
($\SD$ and $\KV$) of net-proton distributions. Those are Au+Au
collisions at {\sNN} = 7.7, 11.5, 39, 62.4 (year 2004) and 200 GeV
(year 2004), Cu+Cu collisions at {\sNN} = 22.4, 62.4, 200 GeV,
$d$+Au at {\sNN} = 200 GeV (year 2003) and $p$+$p$ collisions at
{\sNN} = 62.4 (year 2006), 200 GeV (year 2009). The results for
Au+Au collisions at {\sNN} = 62.4 and 200 GeV have been published in
the paper~\cite{PRL}. The errors shown in the figures are
statistical error only.
\begin{figure}[htb]
 \hspace{-2cm}
\begin{minipage}[t]{0.6\linewidth}
\centering \vspace{0pt}
    \includegraphics[scale=0.19]{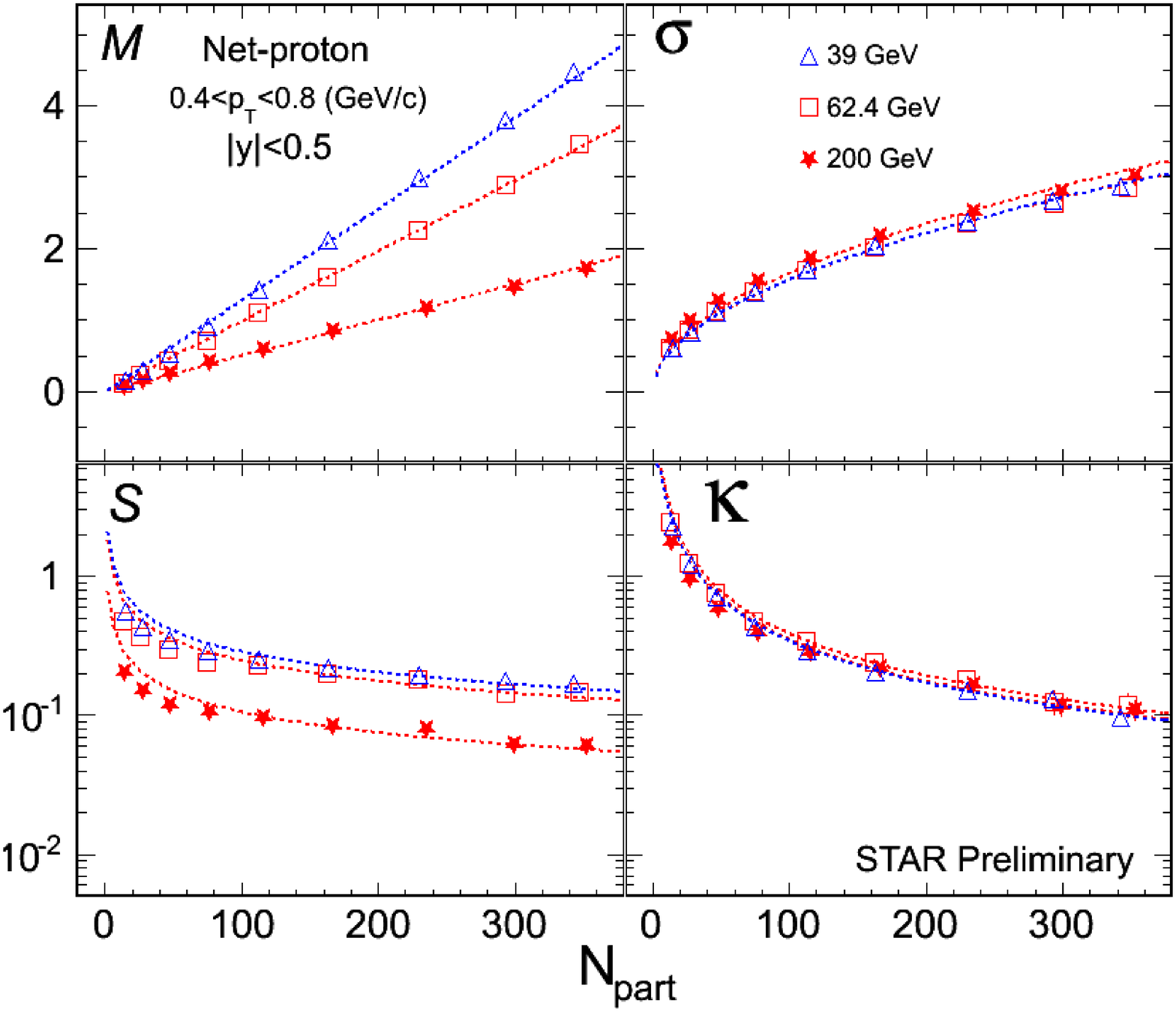}
   \caption{Centrality dependence of various moments of net-proton
multiplicity distributions for Au+Au collisions at {\sNN} =39, 62.4,
200 {\gev}. The dashed lines shown in the figure are expectation
lines from CLT.} \label{fig:Scaling_AuAu}
  \end{minipage}%
  \hspace{0.3in}
  \begin{minipage}[t]{0.6\linewidth}
  \centering \vspace{0pt}
   \includegraphics[scale=0.34]{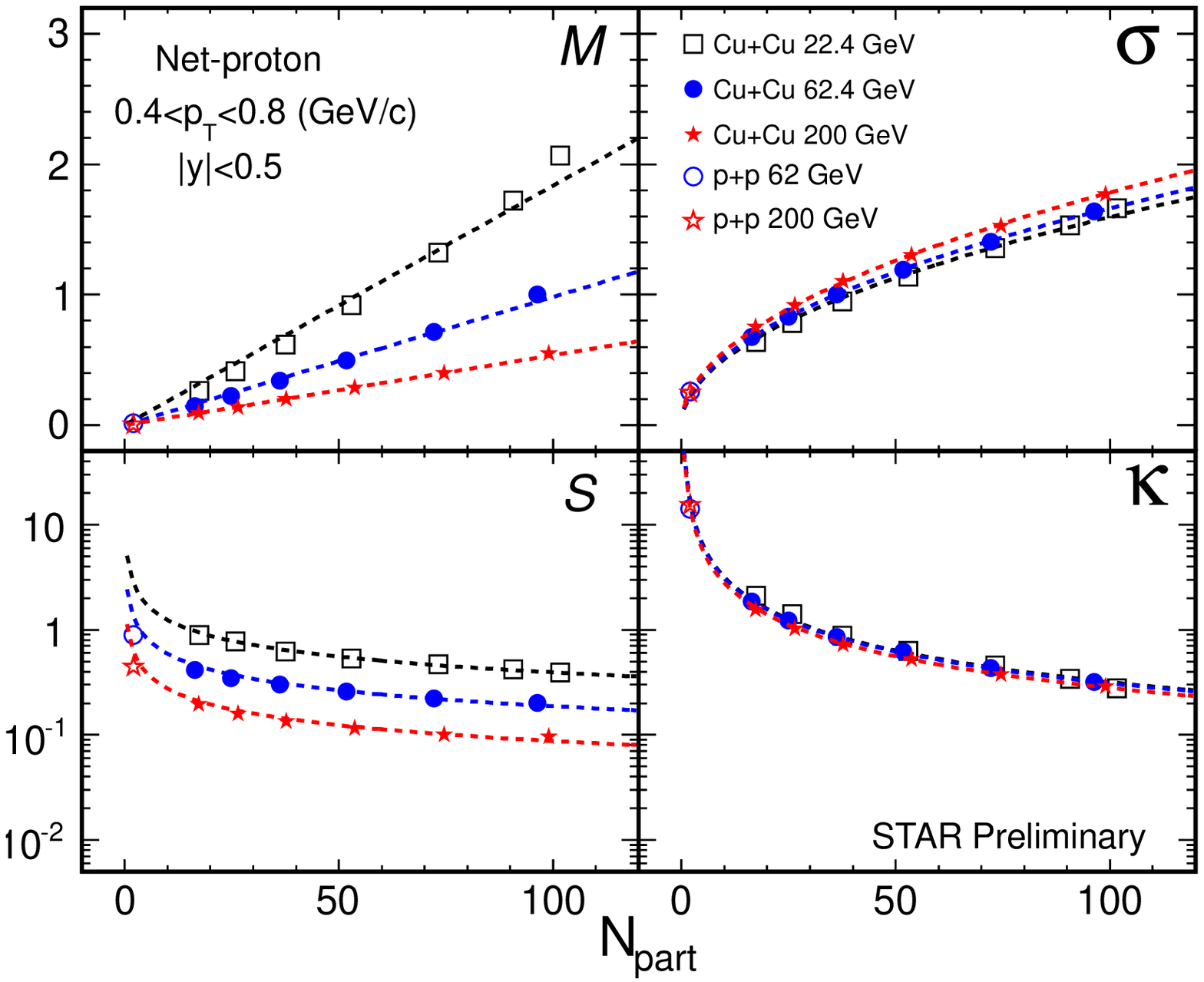}
    \caption[]
    {Centrality dependence of various moments of
net-proton distributions for Cu+Cu collisions at {\sNN} = 22.4, 62.4
and 200 {\gev} and $p$+$p$ collisions at {\sNN} = 62.4 and 200
{\gev}. The dashed lines shown in the figure are expectation lines
from CLT.} \label{fig:Scaling_CuCu_pp}
  \end{minipage} %
\end{figure}
Centrality dependence of various moments of net-proton distributions
for Au+Au, Cu+Cu and $p$+$p$ collisions are shown in Fig.
\ref{fig:Scaling_AuAu} and Fig. \ref{fig:Scaling_CuCu_pp},
respectively. The $M$ and $\sigma$ are found to be monotonically
increasing with increasing of $N_{part}$, while the $S$ and $\kappa$
are decreasing. The various moments can be well described by the
dashed lines shown in the figures, which are derived from Central
Limit Theorem (CLT) by assuming the colliding system consists of
many identical and independent emission
sources~\cite{WWND2011,SQM2009}.
\begin{figure}[htb]
\begin{center}
    \includegraphics[scale=0.45]{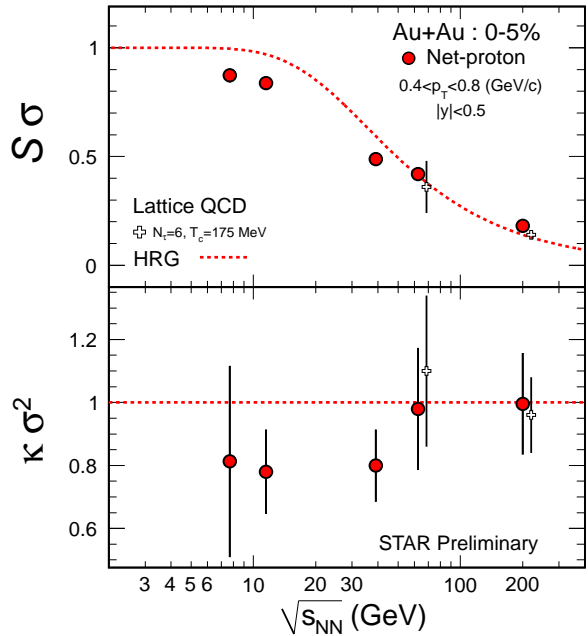}
    \caption[{\KV} of net-proton distributions as a function of Nsigma proton ( $Z_{p}$ )]
    {Energy dependence of moment products ({\KV} and {\SD})
of net-proton distributions for 0-5\% most central Au+Au collisions.
The red dashed lines denote the HRG model calculations, and the
empty markers denote Lattice QCD results~\cite{science}. }
\label{fig:SD_KV_energy}
\end{center}
  \end{figure}
Energy dependence of {\SD} and {\KV} for 0-5\% most central Au+Au
collisions are shown in Fig. \ref{fig:SD_KV_energy}. We find that
the data are consistent with Lattice QCD and HRG model calculations
at high energies ({\sNN} = 62.4 and 200 GeV), while deviations from
HRG model are observed at low energies ({\sNN} = 7.7, 11.5 and 39
GeV). The possible reasons for the deviations are discussed in
~\cite{Neg_Kurtosis,chiral_HRG}.

\section{Summary and Outlook}
Higher moments of net-proton distributions are applied to search for
the QCD critical point and probe the bulk properties of QCD matters.
In summary, we present the measurements of higher moments of
net-proton distributions for Au+Au, Cu+Cu, $d$+Au and $p$+$p$
collisions from STAR experiment. The moment products {\KV} and {\SD}
of net-proton distributions from 0-5\% most central Au+Au collisions
are consistent with Lattice QCD and HRG model calculations at high
energies (\sNN\ = 62.4 and 200 GeV), while the results are smaller
than HRG model calculations at low energies (\sNN\ = 7.7, 11.5, 39
GeV). The analysis of data from another two energies at {\sNN} =
19.6 and 27 GeV, which were collected in the year 2011, are ongoing.

\section{Acknowledgement}
The work was supported in part by the National Natural Science
Foundation of China under grant No. 11135011.

\bibliography{SQM2011}
\bibliographystyle{unsrt}
\end{document}